\documentclass[%
reprint,
superscriptaddress,
 amsmath,amssymb,
prb,
]{revtex4-1}

\usepackage{graphicx}
\usepackage{dcolumn}% Align table columns on decimal point
\usepackage{bm}% bold math
\usepackage{multirow}
\usepackage{mathptmx}
\usepackage{siunitx}
\usepackage{color}

\newcommand{\beq}[1]{\begin{equation}\label{#1}}
\newcommand{\eep}{\;.\end{equation}}
\newcommand{\eec}{\;,\end{equation}}
\newcommand{\eeq}{\end{equation}}

\newcommand{\lb}{\left(}
\newcommand{\rb}{\right)}

\renewcommand{\a}{\alpha}
\renewcommand{\b}{\beta}

\newcommand{\p}{\phi}
\renewcommand{\S}{\Sigma}

\newcommand{\eo}{{\epsilon}_0}

\newcommand{\x}{\chi}
\newcommand{\xn}{\chi_{\eta}}

\newcommand{\z}{\xi}

\newcommand{\ka}{\kappa_a}
\newcommand{\kc}{\kappa_c}
\newcommand{\ks}{\kappa_s}

\newcommand{\pI}{\phi_{\text{I}}} % Potentials
\newcommand{\pII}{\phi_{\text{II}}}
\newcommand{\pIII}{\phi_{\text{III}}}

\newcommand*\dd{\mathop{}\!\mathrm{d}} %differential d
\newcommand\del{\partial} % partial d
\newcommand{\grad}{\nabla} % Grad

\newcommand{\bigO}{\mathcal{O}} % Big O notation

\newcommand*\chem[1]{\ensuremath{\mathrm{#1}}} % for chemical symbols

\newcommand{\F}{\mathcal{F}} % Free energy

\DeclareMathAlphabet{\mathcal}{OMS}{cmsy}{m}{n} % Changes font for mathcal but leaves the rest of the math fonts in Times.

\newcommand{\dm}{d_{\text{m}}}
\newcommand{\dc}{d_{\text{c}}}
\newcommand{\dinf}{d_{\infty}}
\newcommand{\acrit}{\alpha_{\text{c}}}

\begin{document}

%%% TITLE, AUTHORS, ABSTRACT%%%

%\title{Electrostatic Environmental Effects on the Domain Structure of Ferroelectric Thin Films and Superlattices}% Force line breaks with \\

\title{Electrostatics and domains in ferroelectric superlattices}

\author{Daniel Bennett}
 \affiliation{Theory of Condensed Matter, Cavendish Laboratory, Department of Physics, J J Thomson Avenue, Cambridge CB3 0HE, United Kingdom}
 
\author{Maitane Mu\~{n}oz Basagoiti}
\affiliation{Faculty of Science and Technology, University of the Basque Country, Barrio Sarriena 48940 Leioa, Spain}
\affiliation{Gulliver Lab UMR 7083, ESPCI PSL Research University, 75005 Paris, France}
\affiliation{CIC Nanogune and DIPC, Tolosa Hiribidea 76, 20018 San Sebastian, Spain}

\author{Emilio Artacho}
\affiliation{Theory of Condensed Matter, Cavendish Laboratory, Department of Physics, J J Thomson Avenue, Cambridge CB3 0HE, United Kingdom}
\affiliation{CIC Nanogune and DIPC, Tolosa Hiribidea 76, 20018 San Sebastian, Spain}
\affiliation{Ikerbasque, Basque Foundation for Science, 48011 Bilbao, Spain}
%\cortext[cor1]{db729@cam.ac.uk}

\date{\today}

\begin{abstract}
  The electrostatics arising in ferroelectric/dielectric two-dimensional 
heterostructures and superlatitices is revisited here within a simplest Kittel model, 
in order to define a clear paradigmatic reference for domain formation.
  The screening of the depolarizing field in isolated ferroelectric or polar 
thin films via the formation of 180$^{\circ}$ domains is well understood, whereby
the width of the domains $w$ grows as the square-root of the film thickness $d$, 
following Kittel's law, for thick enough films ($w\ll d$).
  This behavior is qualitatively unaltered when the film is deposited on a dielectric
substrate, sandwiched between dielectrics, and even in a superlattice setting, with
just a suitable renormalisation of Kittel's length.
  As $d$ decreases, $w(d)$ deviates from Kittel's law, reaching a minimum 
and then diverging onto the mono-domain limit for thin enough films, always 
assuming a given spontaneous polarization $P$ of the ferrolectric, only modified by
linear response to the depolarizing field.
  In most cases of experimental relevance $P$ would vanish before reaching 
that thin-film regime. 
  This is not the case for superlattices. 
  Unlike single films, for which the increase of the dielectric constant of 
the surrounding medium pushes the deviation from the Kittel's regime to
lower values of $d$, there is a critical value of the relative thickness 
of ferroelectric/dielectric films in superlattices beyond which 
that behavior is reversed, and which defines the
separation between strong and weak ferroelectric coupling in
superlattices.
%  \textcolor{blue}{The different dielectric environments for a ferroelectric thin film
%are presented in detail here for comparison with the superlattice case,
%including all the electrostatic expressions and derivations, which are 
%quite scattered in the literature.}
\end{abstract}

\pacs{Valid PACS appear here}% PACS, the Physics and Astronomy
\maketitle

%%% MAIN TEXT %%%

\section{Introduction}

  The formation of ferromagnetic and ferroelectric domain structures in thin 
films is a well-known phenomenon\cite{kittel,kmf}. 
  Polydomain structures appear in ferroelectric thin films in order to screen the 
%polar discontinuities that occur
electric depolarizing field arising
at the interfaces between the surfaces of the thin film and its environment, 
such as vacuum or a non-metallic substrate. 
  The electrostatic description of a ferroelectric thin film in an infinite vacuum 
has been studied in detail \cite{springer,vacuum_1}. 
%For larger thickness, 
The equilibrium domain width $w$ follows Kittel's law versus film thickness $d$,
namely,  $w\propto \sqrt{d}$, when $w\ll d$. 
  Within the same model but making no approximations on the electrostatics
arising from an ideal, regular polydomain structure, for $w\gtrsim d$, $w$ reaches a 
minimum and grows again when decreasing $d$, until the monodomain is 
reached.\cite{springer,vacuum_1}. 
  This description of an isolated thin film, however,  does not describe the effect 
that surrounding materials have on the thin film and hence the domain structure.

\begin{figure}[!t]
%\hspace*{0.2cm}
\centering
\includegraphics[width=\columnwidth]{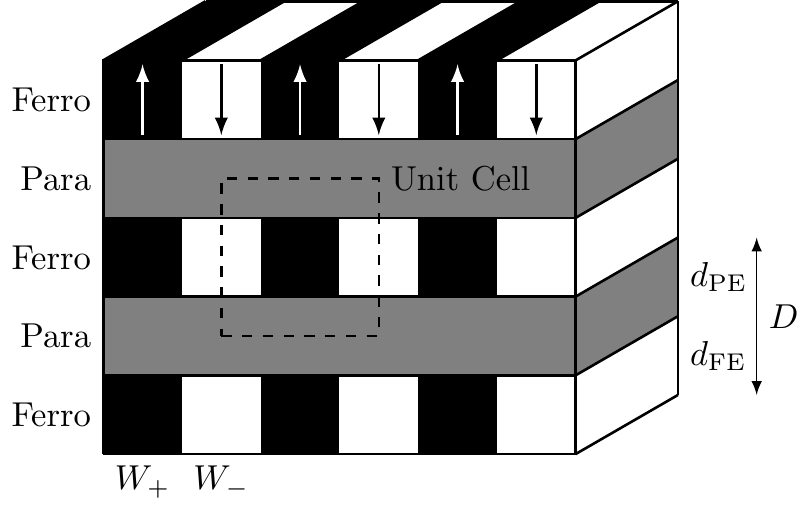}
\caption{Geometry of a FE/PE periodic superlattice. The unit cell is indicated by the dashed square. The thicknesses of the layers are indicated on the right and $W_+$ and $W_-$ are the widths of the positive and negative polarization domains. In polydomain limit, these widths are equal: $W_+ = W_- = w$.}
\label{fig:superlattice}
\end{figure}

It is now possible to fabricate ferromagnetic and ferroelectric samples by growing alternating layers of different thin films, just a few unit cells in thickness, in a periodic array (superlattice)\cite{sl_magnet,sl_bto_sto,sl_bto_sto_2}. Alternating between ferroelectric and paraelectric layers (FE/PE superlattice, see Fig. \ref{fig:superlattice}), a great deal of control over the superlattice's properties can be achieved by changing the relative thicknesses of the layers\cite{dawber_superlattice_1,dawber_superlattice_2,dawber_superlattice_3,dawber_superlattice_4}. This has generated interest in the study of FE/PE superlattices from the theoretical\cite{russian_domains_1,russian_domains_2} and computational\cite{zubko_superlattice_1} perspectives.

The dependence of the domain structure on superlattice geometry cannot be described using the theory of a thin film in an infinite vacuum, however. Some generalizations have appeared in the literature which include the effects of surrounding materials\cite{substrate_domains_1,substrate_domains_1,superlattice_domains_1,superlattice_domains_2,superlattice_domains_3,
superlattice_domains_4,superlattice_domains_5,superlattice_domains_6,russian_domains_1}. For a free-standing thin film on a substrate, it was claimed that the electrostatic description is the same as for a thin film of half the width sandwiched between two paraelectric media \cite{substrate_domains_1}. This has been used to fit measurements of ferroelectric domains \cite{substrate_domains_2,substrate_domains_3}.

By placing a ferroelectric thin film together with a paraelectric layer between two short circuited capacitor plates, it was found that the domain structure could be controlled by tuning the properties of the paraelectric layer, and the stability of the ferroelectric film could be improved \cite{superlattice_domains_1,superlattice_domains_2,superlattice_domains_3,
superlattice_domains_4,superlattice_domains_5,superlattice_domains_6}. This system is to some extent equivalent to a FE/PE superlattice since the capacitor plates impose periodic electrostatic boundary conditions.

  A number of experimental and computational advances have revived interest in this 
%well-known
problem. 
  Interesting effects can occur at interfaces between different materials 
such as the formation of a two-dimensional electron gas (2DEG) at polar-nonpolar 
interfaces like \chem{LaAlO_3/SrTiO_3} (LAO/STO)\cite{ohtomo2002,ohtomo2004}. 
  It is thought that the 2DEG appears to screen the polar discontinuity at the LAO/STO interface\cite{bristowe_electronic_reconstruction}, 
and similarly, it has recently been proposed as a mechanism to screen the depolarising field at ferroelectric/paraelectric interfaces\cite{pablo_2deg_theory,pablo_2deg_dft}. 
  This is difficult to observe directly by experiments, and evidence for 2DEG formation at FE/PE interfaces has only very recently been found\cite{2deg_2018,2deg_2018_2,2deg_2018_3}. 
  This is because there is competition with domain formation for the screening of the 
depolarizing field.
  Since these phenomena are of an electrostatic origin, a clear picture of the electrostatics of ferroelectrics is essential in order to understand them.

  Although ferroelectric thin films have been frequently simulated from first 
principles in different settings and environments\cite{pablo_2deg_dft,
junquera_critical_thickness,junquera_critical_thickness_2,dft_1,dft_2,dft_3},
ferroelectric domains are quite demanding to simulate from first principles,
as they require much larger supercells. 
  Recent developments in effective modelling from first-principles calculations 
(second-principles methods) make it possible to study very large systems, 
including large domain structures in ferroelectric materials\cite{vanderbilt_bto_effective_prl,vanderbilt_bto_effective_prl_2,
vanderbilt_bto_effective_prb,joannopoulos,2nd_principles_theory,2nd_principles_theory_2,bellaiche_1,bellaiche_2,bellaiche_3}
and observe interesting related effects such as negative capacitance\cite{zubko_negative_capacitance} and polar skyrmions\cite{2nd_principles_skyrmion}.
  These scientific advances, both experimental and computational, have motivated 
us to revisit the electrostatic description of ferroelectric domains.

\begin{figure}[!t]
%\hspace*{0.4cm}
\centering
\includegraphics[width=\columnwidth]{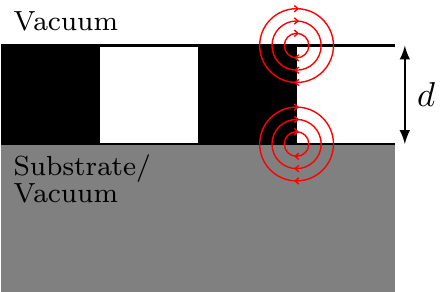}
\caption{Geometry of a ferroelectric thin film of thickness $d$ with a $180^{\circ}$ polydomain structure. The red lines represent the electrostatic depolarizing field, which bend around the interfaces and domain walls.}
\label{fig:film}
\end{figure}

The continuum electrostatic description of a monodomain ferroelectric thin film is essentially unaffected by a dielectric environment of the film. This is because there is zero field outside the thin film and hence these regions make no contributions to the electrostatic energy. For a polydomain ferroelectric thin film, the domain structure introduces stray electric fields into the regions outside the film (see Fig. \ref{fig:film}). We would expect different behavior if we replaced the vacuum regions with a dielectric medium. Understanding the effect of more general geometries on the electrostatic description of ferroelectric thin films not only gives an insight into how the surrounding dielectric media contribute to the screening of the depolarizing field, but also allows us to understand the behavior of the domain structure of the film in different environments, bringing us closer to a realistic description of a thin film.

This paper is organized as follows: first, we review the continuum model of a ferroelectric thin film in a vacuum with full electrostatics and a domain wall term. We then generalize the theory for three different systems: a thin film on an infinite substrate (overlayer, OL), a thin film sandwiched between two infinite dielectric media (sandwich, SW), and a FE/PE superlattice (SL). 
We keep the prevalent nomenclature in the literature of referring to a spacer material such
as STO as paraelectric, but the description will be exclusively that of a dielectric 
material with a given isotropic dielectric permitivity.

Some of these systems have appeared in the literature in various contexts and with different levels of detail. Here we present a coherent comparative study. We compare the different cases in the Kittel limit first, when $w\ll d $, and for which analytic expressions are obtained for $w(d)$. In the 
general situation, and, in particular, when $w \gtrsim d$, the behavior of the domain width must be determined numerically. Previous studies of periodic superlattices have assumed ferroelectric and paraelectric layers of equal width. Here we provide a more general study of domain structures as a function of superlattice geometry, and obtain an interesting critical value for their thickness ratio
which separates the weak and strong coupling regimes in SL structures. We also present a detailed derivation of the electrostatic energies in Appendix A.

\section{Review of model for a film in vacuum}

  The fundamental model used in this work is based on 
the following free energy (per unit volume) of a ferroelectric thin film in a 
vacuum with a $180^{\circ}$ stripe domain structure 
\beq{F_tot}
\F = \F_0(P) + \frac{\S}{w} + \F_{\text{dep}}(w,d) \, ,
\eec
where $\F_0(P)$, defined as
\beq{eq:ferro}
\F_0(P) = \frac{1}{2\eo\kappa_c}\lb \frac{1}{4}\frac{P^4}{P_S^2}-\frac{1}{2}P^2\rb \, ,
\eeq
is the bulk ferroelectric with spontaneous polarization $P_S$ and dielectric permittivity $\kc$, which describes the curvature about the minima. $\S$ is the energy cost (per unit area) for creating a domain wall, $\F_{\text{dep}}$ is the electrostatic energy associated with the depolarizing field, and $w$ and $d$ are the width of one domain and thickness of the film, respectively. We assume that the domain walls are infinitely thin. We will consider the polarization oriented normal to the FE film. Furthermore, we assume that the polarization is constant throughout this film. In reality, the magnitude of the polarization will increase or decrease near the interfaces due to surface effects. This effect can be treated using a Landau-Ginzburg theory, but will be neglected here.

  Since we will be interested in the electrostatic effects due to a finite polarization, we will
consider the polarization to be $P_S$, except for its modification in linear response to
the depolarizing field implicit when using a dielectric permitivity for the material normal
to the field, $\kappa_c$. 
  This assumption is equivalent to replacing the form of $\F_0(P)$ in Eq.~\ref{eq:ferro} by
\beq{eq:ferro2}
\F_0(P) = \frac{1}{2\eo\kappa_c} ( P-P_S)^2  \, .
\eeq
The equilibrium domain structure for this system for a given thickness is obtained by minimising the energy: $\partial_{w}\F=0$. 

\begin{figure}[t] % [!h] 
\centering
%\hspace*{-1cm}
\includegraphics[width=\columnwidth]{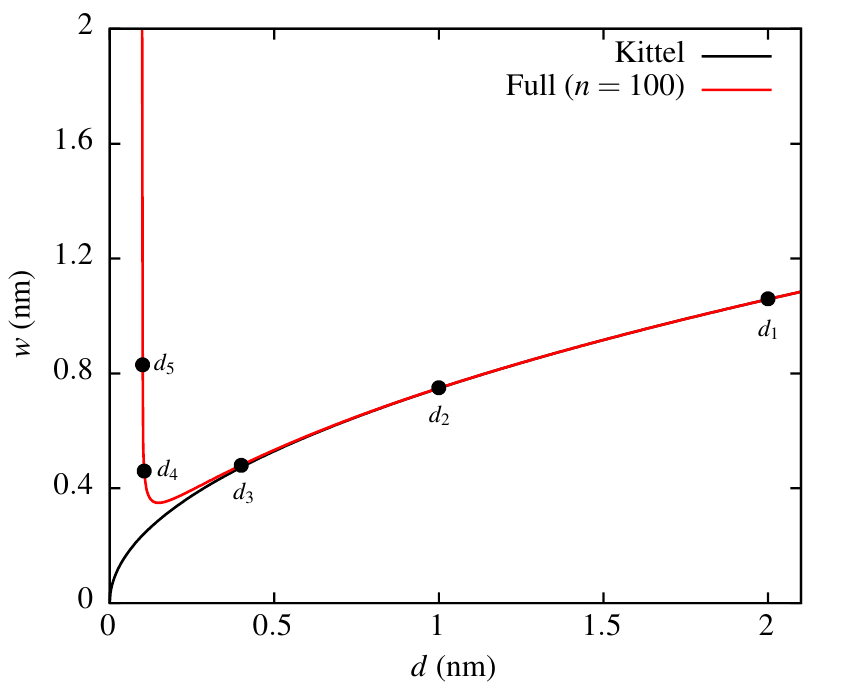}
\caption[region1]{Equilibrium domain width as a function of thickness for an isolated thin film. The red curve shows the numerical solution using the full expression for the electrostatic energy, truncated at $n=100$ terms. The solid black curve is the Kittel curve: $w(d) = \sqrt{l_k d}$. Some of the points $(d_i,w_{eq,i})$ are marked with black dots, which will be referred to in Fig. \ref{fig:vacuum_energy}. The following values of $d$ were used: $d_1 = 2 \ \si{nm}$, $d_2 = 1 \ \si{nm}$, $d_3 = 0.4 \ \si{nm}$, $d_4 = 0.105 \ \si{nm}$, $d_5 = 0.1 \ \si{nm}$, $d_6 = 0.99 \ \si{nm}$, $d_6 = 0.9 \ \si{nm}$. The values of the parameters used are: $P = 0.78$ C/m\textsuperscript2, $\S = 0.13$ J/m\textsuperscript2, $\ka = 185$, $\kc = 34$, $\ks = 300$.}
\label{fig:vacuum_domain}
\end{figure}

In this work we consider an ideal domain structure made by regular straight stripes,
all of them of the same width $w$ (in Appendix A different widths are considered).
For an isolated film, the electrostatic energy for that structure is given by \cite{springer}
\beq{vacuum_full}
\F_{\text{e}} = \frac{8P_S^2}{\eo\pi^3}\frac{w}{d}\sum_{n \ \text{odd}}\frac{1}{n^3}\frac{1}{1+\x\kc\coth{\lb \frac{n\pi}{2}\x\frac{d}{w}\rb}}
\eeq
where $\ka$, $\kc$ are the dielectric permittivities in the directions parallel and normal to the film and $\x=\sqrt{\ka/\kc}$ is the anisotropy of the film. In the Kittel limit\cite{kittel,kmf}, $\frac{w}{d}\ll 1$, Eq. \eqref{vacuum_full} reduces to
\beq{F_kittel}
\F^{\text{K}}_{\text{e}} = \frac{P_S^2}{2\eo}\b\frac{w}{d}
\eec
where
\beq{beta}
\b = \frac{14\zeta(3)}{\pi^3}\frac{1}{1+\x\kc}
\eec
and $\zeta(n)$ is the Riemann zeta function. An analytic expression is obtained for the equilibrium domain width:
\beq{kittel_vacuum}
w(d) = \sqrt{l_kd}
\eec
where
\beq{}
l_k = \frac{2\eo\S}{P_S^2\b}
\eeq
is the Kittel length, which defines a characteristic length scale of the system. Eq. \eqref{kittel_vacuum} is known as Kittel's law\cite{kittel}.

  Beyond the Kittel regime, we can obtain the equilibrium domain width from 
the numerical solution to Eq. \eqref{F_tot} for the full electrostatics expression
in Eq.~\eqref{vacuum_full}. 
  In Fig. \ref{fig:vacuum_domain}, we plot $w(d)$ both from the Kittel limit and numerical solutions, truncating Eq. \ref{vacuum_full} at $n=100$ terms. We use \chem{PbTiO_3} (PTO) and \chem{SrTiO_3} (STO) as examples of ferroelectric and paraelectric materials, respectively, in all of the plots in this paper, using suitable parameters\footnote{The following values of $d$ were used: $d_1 = 2 \ \si{nm}$, $d_2 = 1 \ \si{nm}$, $d_3 = 0.4 \ \si{nm}$, $d_4 = 0.105 \ \si{nm}$, $d_5 = 0.1 \ \si{nm}$, $d_6 = 0.99 \ \si{nm}$, $d_6 = 0.9 \ \si{nm}$. The values of the parameters used are: $P = 0.78$ C/m\textsuperscript2, $\S = 0.13$ J/m\textsuperscript2, $\xn = 26$, $\ka = 185$, $\kc = 34$, $\ks = 300$.}. From Fig. \ref{fig:vacuum_domain} we see that the domain width follows Kittel's law at large values of $d$, but, for decreasing $d$, $w$ reaches a minimum at $\dm$ and then 
diverges at $d_{\infty}$.

\begin{figure}[b] % [!h] 
\centering
%\hspace*{-1cm}
\includegraphics[width=\columnwidth]{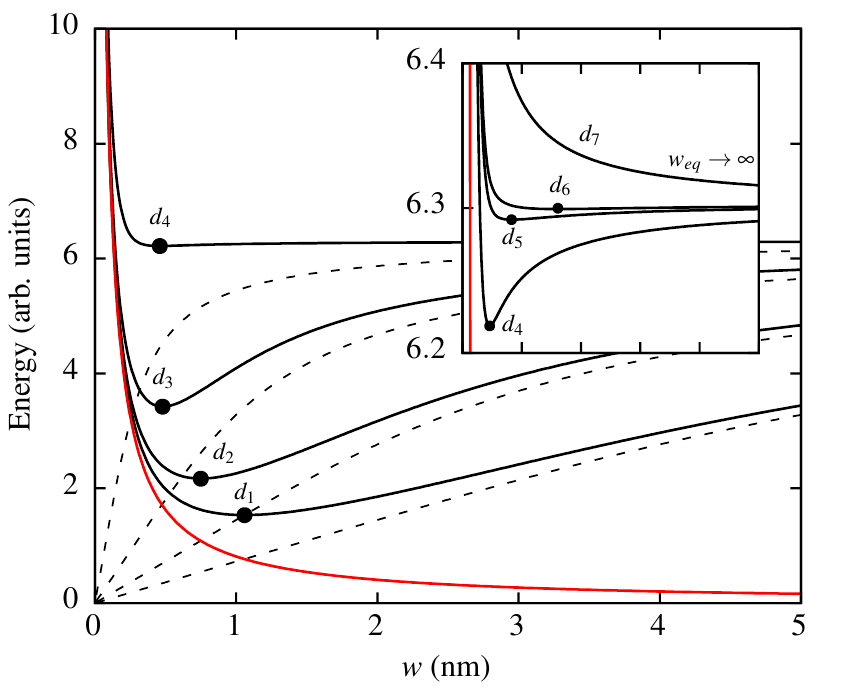}
\caption[region1]{Energy as a function of $w$ for various values of $d$. The red curve is the energy cost of creating domain walls. The black curves are the total energies for different values of $d$, and the dashed curves immediately beneath are the respective electrostatic energies at the same thicknesses (truncated at $n=100$ terms). The minimum with respect to $w$ is indicated with a black dot. The inset shows the energy curves near where the equilibrium domain width diverges.}
\label{fig:vacuum_energy}
\end{figure}

% While Eq. \eqref{vacuum_full} is complicated, and prevents us from obtaining analytic solutions, 
We can understand this behavior by studying the shape of the energy curves as a function of $w$ and $d$. In Fig. \ref{fig:vacuum_energy}, we plot the different energies as a function of domain width at different thicknesses. The energy per unit volume associated with creating the domain walls, shown in red, is unaffected by the thickness of the film. The dashed gray lines show the electrostatic energy Eq. \eqref{vacuum_full} at different thicknesses. We can see in each case that for small $w$, they are linear in $w$, following Kittel's law (Eq. \eqref{F_kittel}). As $w$ increases, Kittel's law breaks down, and the curves level off, approaching the monodomain electrostatic energy:
\beq{F_mono}
\F_{\text{mono}} = \frac{P_S^2}{2\eo\kc}
\eep
As $d$ decreases, the saturation of the electrostatic energy is realized earlier, and the minimum in  total energy becomes shallower, eventually disappearing, the equilibrium domain width thereby diverging. We can visualize this by looking at the minima of the total energy curves as $d$ is decreased. The minima are marked with black dots on Fig. \ref{fig:vacuum_energy} and are also shown on the plot of $w(d)$ in Fig. \ref{fig:vacuum_domain}.

\begin{figure}[t] % [!h] 
\centering
%\hspace*{-1cm}
\includegraphics[width=\columnwidth]{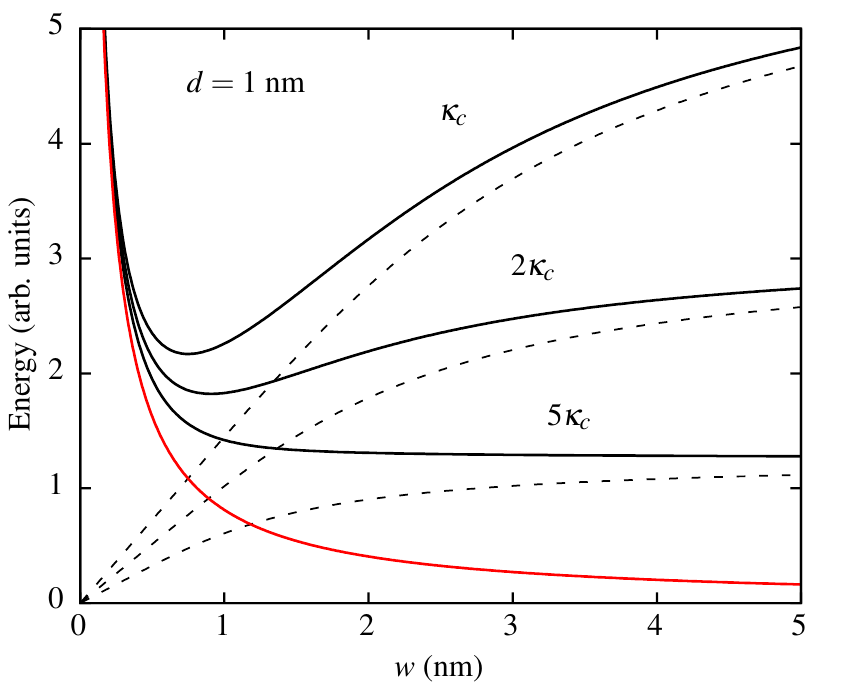}
\caption[region1]{Energy vs $w$ for various multiples of $\kc = 34$. The red curve is the energy cost of creating a domain structure. The black curves are the total energies for different values of $d$, and the dashed grey curves immediately beneath are the respective electrostatic energies at the same thicknesses (truncated at $n=100$ terms). Each curve has a thickness of $d = 1 \ \si{nm}$. .}
\label{fig:vacuum_kc}
\end{figure}

The described deviation from Kittel's law is sensitive to the system's parameters. In Ref. [\onlinecite{vacuum_1}], an expression for $d_m$ was reported\footnote{The authors in Ref. [\onlinecite{vacuum_1}] do not provide details on how this was obtained.} of the form
\beq{dcrit}
\dm = 5\pi\S\eo\frac{\kc}{\x}\frac{1}{P_S^2}
\eep
where such dependence is explicit.

 In Fig. \ref{fig:vacuum_kc} we show the effect of changing $\kc$. Increasing $\kc$ decreases the curvature of the electrostatic energy and also decreases the monodomain energy (the asymptotic energy for large $w$). By increasing $\kc$ for a fixed value of $d$, the total energy minimum again
becomes shallower and then disappears. 

Although analytic solutions for the equilibrium domain width can not be obtained using Eq. \eqref{vacuum_full}, we can obtain approximate solutions. Close to the thickness where the domain width diverges, $\dinf$, we have
\beq{w_asymptote}
\begin{split}
w(d) &\cong \frac{\pi\x}{2\sqrt{e}} d \exp{\lb\frac{\pi^2}{8}\frac{\kc}{\x}\b\frac{l_k}{d}\rb}\\
\dm &\cong \frac{\pi^2}{8}\frac{\kc}{\x}\b l_k = \frac{\pi^2}{4}\S\eo\frac{\kc}{\x}\frac{1}{P^2}
\end{split}
\eep
Details of this approximation are given in Appendix B and in Ref. [\onlinecite{superlattice_domains_1}]. 
  In this approximation $d_m$ has the same dependence on the systems parameters 
as Eq. \eqref{dcrit}, but the constant prefactor is different.

  We can also obtain an analytic approximation to the domain width at all thicknesses by replacing the Eq. \eqref{vacuum_full} with a simpler expression which has the correct behavior in the monodomain and Kittel limits,
\beq{F_approx}
\F_{\text{e}}^* = \frac{P_S^2}{2\eo\kc}\frac{1}{1+\frac{1}{\kc\b}\frac{d}{w}}
\eec
which clearly tends to Eq. \eqref{F_mono} and Eq. \eqref{F_kittel} when $w/d$ is large and small, respectively. Using this, we get
\beq{w_approx}
\begin{split}
w(d) &= \frac{\sqrt{l_k d}}{1-\kc\b\sqrt{\frac{l_k}{d}}}\\
\dm &= 4\kc^2\b^2l_k(\ks) \approx \frac{112\zeta(3)}{\pi^3}\S\eo\frac{\kc}{\x}\frac{1}{P_S^2}
\end{split}
\eep
Details of this approximation are given in Appendix C. This approximation is of the same form as Eq. \eqref{dcrit} but again with a different numerical prefactor. Eq. \eqref{w_approx} gives a good approximation to $\dm$, but overestimates the domain width near $\dm$. This is because, while Eq. \eqref{F_approx} has the correct behavior in the monodomain and polydomain limits, it underestimates the curvature in the intermediate region. In spite of this, the approximation predicts the correct dependence on the system's parameters.

Having understood the behavior of the equilibrium domain width with thickness and the system's parameters, we proceed to investigate the effect of changing the surrounding environment of the thin film. In order to investigate the effect of changing the environment, we must obtain more general expressions for the electrostatic energies, similar to Eq. \eqref{vacuum_full}.

\section{Generalized Electrostatics}

The electrostatic energies were obtained for the OL, SW and SL cases. The expressions, including their derivation, are shown in detail in Appendix A.

\subsection{Generalized Kittel Law}

 Taking the Kittel limit for the energies in Eqs. \eqref{full_elec} and \eqref{full_sub}, we obtain a generalization of Kittel's law:

\beq{kittel_general}
\begin{split}
w(d) &=\sqrt{l_k(\ks)d}\\
l_k(\ks)& = \frac{2\eo\S}{P_S^2\b(\ks)}
\end{split}
\eep

The generalization is introduced through the factor $\b$:

\beq{beta_general}
\begin{split}
\b_{\text{SW}}(\ks)& = \frac{14\zeta(3)}{\pi^3}\frac{1}{\ks+\x\kc}\\
\b_{\text{SL}}(\ks,\a) & = \frac{1}{1+\a}\frac{14\zeta(3)}{\pi^3}\frac{1}{\ks+\x\kc}\\
\b_{\text{OL}}(\ks) & = \frac{7\zeta(3)}{\pi^3}\lb \frac{1+\ks +2\x\kc}{(1+\x\kc)(\ks+\x\kc)}\rb
\end{split}
\eep

The SL case has an additional dependence on $\a \equiv d_{\text{PE}}/d_{\text{FE}}$, the ratio of thicknesses of the paraelectric and ferroelectric layers. However, the energy cost of creating a domain wall is also renormalized by this refactor, and thus, in the Kittel limit, the ratio $\a$ affects the energy scale but does not influence the behavior of the domains. For each case in Eq. \eqref{beta_general}, Eq. \eqref{beta} is recovered in the limit $\ks\to 1$.\\

\begin{figure}[t] % [!h]
%\hspace*{0.4cm}
\centering
\includegraphics[width=\columnwidth]{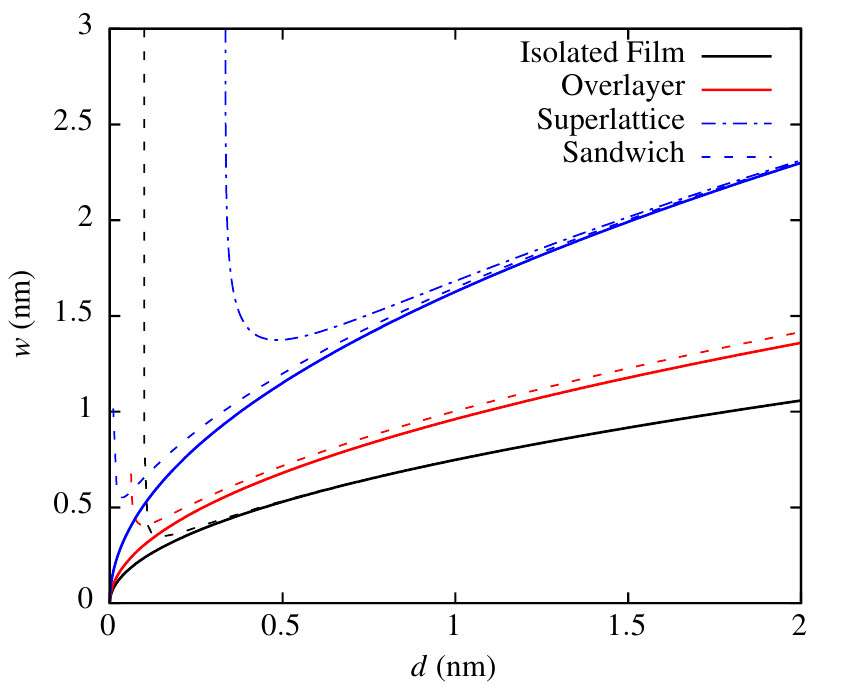}
\caption{$w(d)$ for a thin film in a vacuum (black), the OL system (red), the SW (blue) and the SL system with $\a=3$ (blue). The solid lines show the analytic solutions from the generalized Kittel's law and the dashed lines show numerical solutions using the full expressions for the depolarizing energies. The SL and SW systems have identical square root curves in the Kittel limit. }
\label{fig:kittel}
\end{figure}

The domain widths for the four different systems are plotted in Fig. \ref{fig:kittel}. We can see that including the environment has the effect of shifting the curve upwards, but the square root behavior is unaffected. This makes sense physically: the paraelectric medium contributes to the screening of the depolarizing field. For high dielectric constants, this contribution is large, meaning less screening is required by the domains, so there are fewer domains, and hence the width increases.\\

The SL and SW cases have the exact same behavior in the Kittel limit. This expected, since in the Kittel limit, the field in the superlattice loops in the paraelectric layers but does not penetrate through to neighboring ferroelectric layers. In this regime, the coupling between the ferroelectric layers is weak, and the ferroelectric layers are isolated from each other, which is the SW case.

 In Ref. [\onlinecite{substrate_domains_1}], it was claimed that there should be a factor of two between the length scales of the OL and SW systems. From Eq. \ref{beta_general} we have:

\beq{}
\frac{l_{k,\text{OL}}(\ks)}{l_{k,\text{SW}}(\ks)} = \frac{\b_{\text{SW}}(\ks)}{\b_{\text{OL}}(\ks)} = \frac{1 +\x\kc}{1+\ks +2\x\kc}
\eep

When $\ks\approx 1$, this is indeed true. However, when the $\ks$ is comparable or larger, this does not hold. For example, for PTO and STO, $\x\kc \sim 79$ and $\ks = 300$ and can be as large as $10^4$ at low temperatures, and the difference in the Kittel lengths becomes significantly larger than a factor of two.

\subsection{Beyond Kittel: Thin Films}

\begin{figure}[b] % [!ht]
%\hspace*{0.4cm}
\centering
\includegraphics[width=\columnwidth]{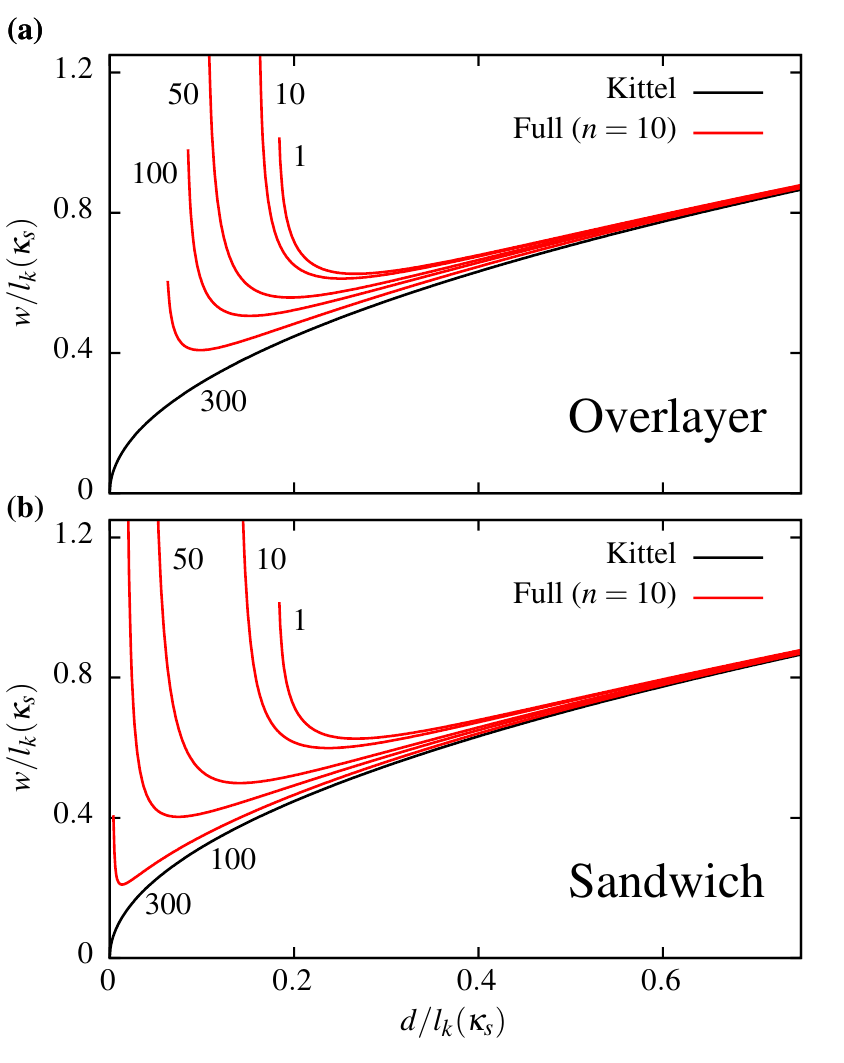}
\caption{Domain widths as a function of thickness for various values of $\ks$ for (a) the OL system and (b) the SW system. Each domain width and film thickness is normalized by the Kittel length for that value of $\ks$. }
\label{fig:ks_sub_dielec}
\end{figure}

\begin{figure}[t] % [!h]
%\hspace*{0.4cm}
\centering
\includegraphics[width=\columnwidth]{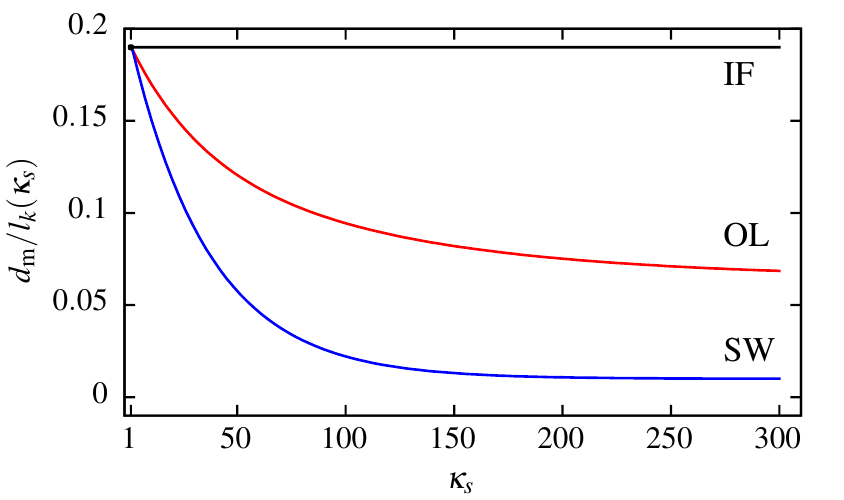}
\caption{$d_m$ relative to the corresponding Kittel length as a function of dielectric permittivity of the substrate material for the OL (red) and the SW (blue) systems. }
\label{fig:sub_ks_min}
\end{figure}

Although the square root curve is simply shifted upwards after including the environment, the behavior for thinner films is quite different. In Fig. \ref{fig:kittel} we can see that the width at which the domain width diverges is very sensitive to the dielectric environment. In Fig. \ref{fig:ks_sub_dielec}, we plot the domain widths for various values of the dielectric permittivity of the substrate material, $\ks$ for the OL and SW systems, each curve scaled by the relevant Kittel length, $\l_k(\ks)$. We see that $\dm$ decreases with increasing $\ks$.
  In Fig. \ref{fig:sub_ks_min} we plot the critical thickness as a function of $\ks$ to illustrate this effect.
For the SW system, $\dm$ decreases more dramatically. This is expected, as there is screening on both sides of the thin film in the SW system.

We can understand the effect of the paraelectric permittivity on $\dm$ by examining the form of the electrostatic energy. For example, for the SW system:
\beq{}
\F_{\text{SW}} = \frac{1}{\ks}\frac{8P_S^2}{\eo\pi^3}\frac{w}{d} \sum_{n\ \text{odd}} \frac{1}{n^3}\frac{1}{1+\x\frac{\kc}{\ks}\coth{\lb \frac{n \pi}{2}\x \frac{d}{w}\rb}}
\eep
This is equivalent to the the electrostatic energy of the IF system, but with the overall energy and $\kc$ both scaled by $\ks$. As we know from Eqs. \eqref{dcrit} and \eqref{w_approx} that $\dm \propto \kc^{3/2}$, it is clear that $\dm$ should decrease with increasing $\ks$.

\begin{figure}[b] % [!h]
\hspace*{-0.4cm}
\centering
\includegraphics[width=\columnwidth]{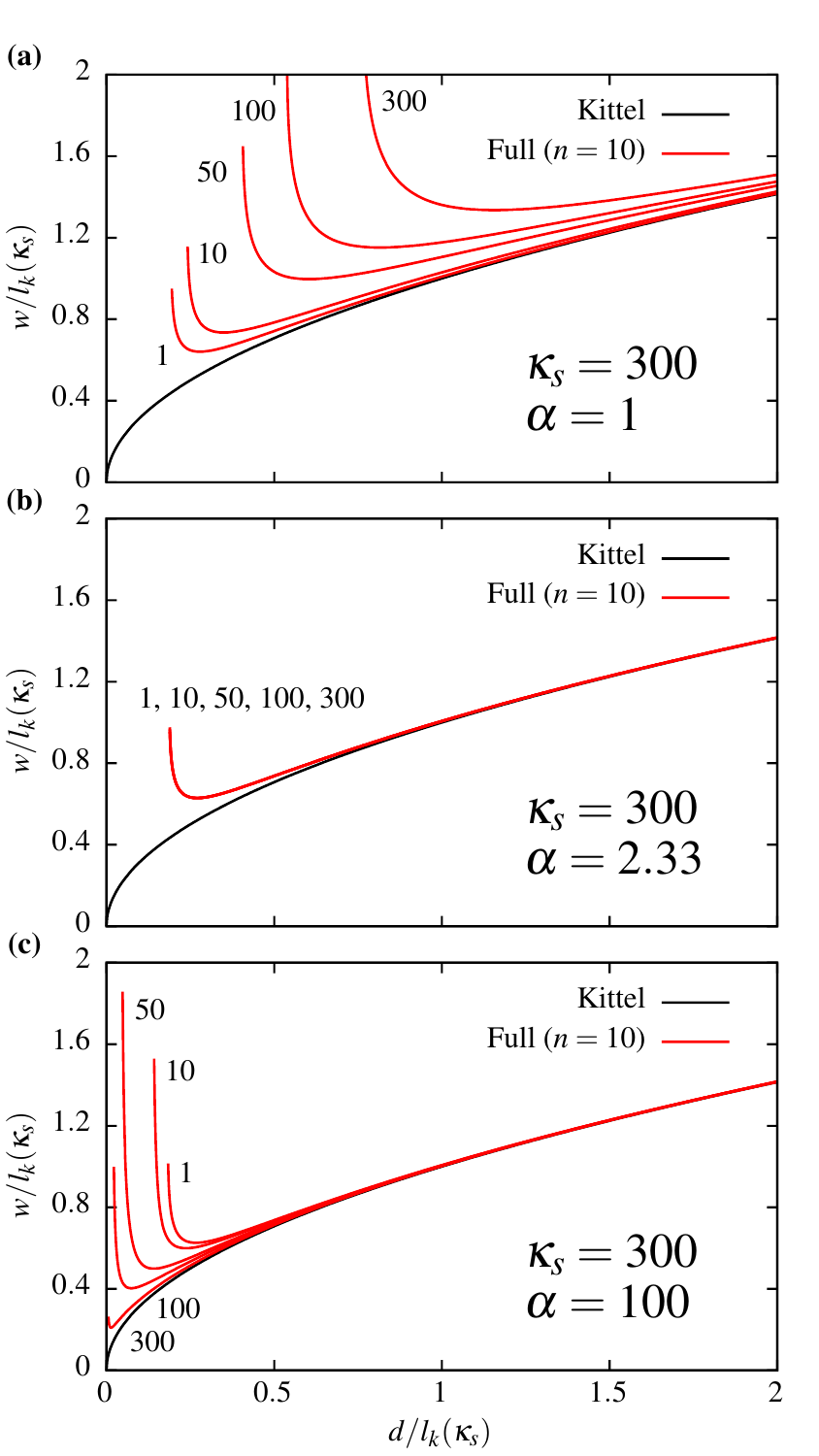}
\caption{Domain width as a function of thickness for the SL system with (a) $\a=1$, (b) $\a=\sqrt{\ka/\kc}=2.33$, and (c) $\a=100$. Each domain width and film thickness is normalized by the Kittel length for that value of $\ks$. }
\label{fig:superlattice_ks}
\end{figure}

\subsection{Superlattice}

For the SL system with with $\a=1$ ($d_{\text{PE}}=d_{\text{FE}}$), we find that $\dm$ actually increases with the permittivity of the paraelectric layers, as shown in Fig. \ref{fig:superlattice_ks}(a),
contrary to what happens for OL and SW.
 For small values of $\a$, the periodic boundary conditions of the superlattice make the electrostatic description very different from the OL and SW systems. When the paraelectric layers are thin, the depolarizing field penetrates through them and there is strong coupling between the ferroelectric layers. The superlattice acts as an effectively uniform ferroelectric material. The average polarization decreases with the permittivity of the paraelectric layers, and according to Eq. \eqref{dcrit},  $d_m$ increases.

For large spacings between the ferroelectric layers ($\alpha \gg 1$), the coupling between them becomes weak, the SW system being realized for $\a\to\infty$. This is illustrated in Fig. \ref{fig:superlattice_ks}(c), which is almost identical to Fig. \ref{fig:ks_sub_dielec}(b).

\begin{figure}[t] % [!h]
%\hspace*{0.4cm}
\centering
\includegraphics[width=\columnwidth]{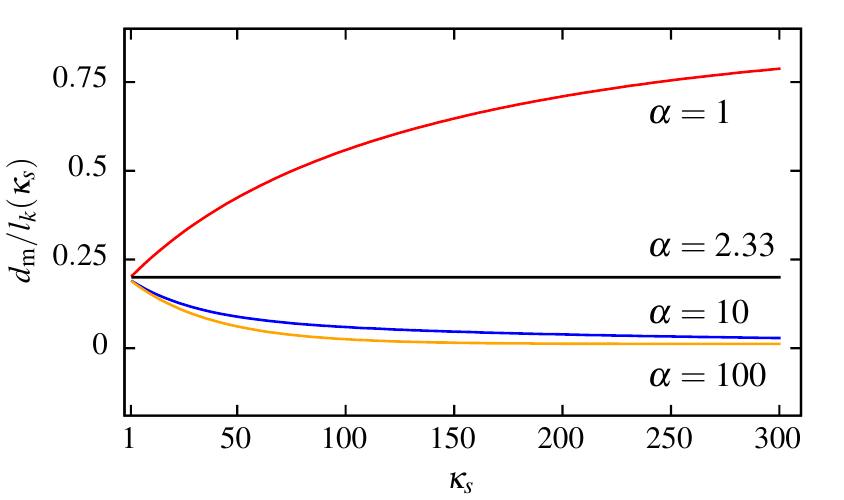}
\caption{Critical thickness of the SL system as a function of $\ks$ for several values of $\a$. Each value of $d$ is scaled by the appropriate Kittel length. }
\label{fig:ks_crit_superlattice}
\end{figure}

Interestingly, when $\a=\acrit\equiv\x \approx 2.33$, $\dm/l_k(\ks)$ is independent of $\ks$. At this ratio, the dielectric permittivity of the spacer has no influence on the equilibrium domain structure, relative to the length scale given by $l_k(\ks)$. This is shown in Fig. \ref{fig:superlattice_ks}(b). In Fig. \ref{fig:ks_crit_superlattice} we plot $d_m$ as a function of $\ks$ for different values of $\a$. We see that when $\a > \acrit$, $\dm$ increases with $\ks$, while it decreases for $\a < \acrit$, and remains constant when $\a=\acrit$. Thus, $\acrit$ represents a natural boundary between the strong and weak coupling regimes of superlattices.

The critical ratio $\acrit$ can be predicted from both the asymptotic and analytic approximations. Using the analytic approximation to the SL system (see Appendix C), we have
\beq{}
\begin{split}
\frac{\dm}{l_k(\ks)} &= 4(\kc+\a^{-1}\ks)^2\b(\ks)^2\\
 & \propto \frac{\kc}{\ka}\frac{(1+\frac{\ks}{\a\kc})^2}{(1+\frac{\ks}{\x\kc})^2}\frac{1}{P_S^2}
\end{split}
\eep
From this we can see that when $\a=\acrit$, the dependence on $\ks$ vanishes.

\section{Discussion and Conclusion}

We have extended the electrostatic description of an isolated ferroelectric thin film within Kittel's  model to thin films surrounded by dieelectric media and FE/PE superlattices. While some of the generalizations have previously appeared in the literature, a detailed comparison had not been done. In doing so, we have understood how the dielectric materials influence the domain structure in the ferroelectric materials, both in the Kittel limit and beyond.

In the Kittel limit, the square root behaviour is only affected in scale, the domain width increasing with dieelectric permittivity, $\ks$. This provides a useful correction to measurements of domain width with film thickness, as Kittel's law for an isolated film typically underestimates domain widths. Beyond Kittel's regime, we have found that increasing $\ks$ decreases $d_m$, that it, the thickness for which the domain width is minimal.

For FE/PE superlattices, we found that $\ks$ can either decrease or increase $d_m$, depending on the ratio of thicknesses, $\a = d_{\text{PE}}/d_{\text{FE}}$. We relate this to the different coupling regimes between the ferroelectric layers, as discussed in Ref. [\onlinecite{russian_domains_1}], for example. When $\a$ is large, the ferroelectric layers are weakly coupled, and the minimum thickness decreases with $\ks$. When $\a$ is small, the ferroelectric layers are strongly coupled, and $d_m$ increases with $\ks$. Remarkably, when $\a = \acrit \equiv \x$, the anisotropy of the ferroelectric layers, $d_m/l_k$ is unaffected by $\ks$. $d_m$ does change, since the Kittel length depends on $\ks$, but the critical ratio $\acrit$ serves as a clear boundary between the strong and weak coupling regimes from an electrostatic viewpoint. %It could be useful for identifying the nature of the coupling in experimentally fabricated superlattice samples, provided the values of $\x$ and $\a$ are known.

This continuum electrostatic theory makes use of several important approximations. Polarization gradients throughout the films occur, as the polarization increases or decreases close to the interfaces, depending on the surface or interface effects, but such gradients have been neglected here. Domains are typically not straight or of infinite length, and the domain structure may not be an equilibrium one ($A\neq 0,\pm1$, see Appendix A). %Different types of domain walls can form, such as $90^{\circ}$ and $109^{\circ}$, which are more difficult to describe.

As stated above, one important approximation in the Kittel-like model used here is the description of the polarization in the ferroelectric, assuming a dielectric linear-response modification of the spontaneous polarization $P_S$ (or using Eq.~\ref{eq:ferro2} instead of Eq.~\ref{eq:ferro} as free energy term related to the polarization). Within this approximation, the system approaches a monodomain phase in a thin-limit regime in which the more complete treatment may predict $P=0$. We investigate this possibility by considering a theory with Eq.~\ref{eq:ferro} for the polarization, and Eq. \ref{F_approx} as the model electrostatic energy. We find that the polarization is zero for 
small thicknesses until
\beq{}
\dc = 27(\kc\b)^2 l_k
\eeq
at which the polarization jumps to $P_S/\sqrt{3}$ \onlinecite{pablo_2deg_theory} and quickly saturates to $P_S$. %The domain width is scaled by $P_S/P$, which can be at most $\sqrt{3}$. 
Or, coming from $d>d_c$, the polarization decreases and the domain width increases, until at $\dc$, the ferroelectric material becomes paraelectric.

If $\dc < d_{\infty}$ the theory is unaffected, and the polydomain to monodomain transition would occur before the ferroelectric to paraelectric transition. Otherwise, the ferroelectric film becomes paraelectric without a polydomain to monodomain transition.  
%A decrease in $P$ is equivalent to increasing $\kc$, which would increase $\dm$ (see Fig. \ref{fig:vacuum_kc}). 
For an isolated thin film of PTO, $\dm \sim 0.2l_k$ and $\dc\sim 0.8l_k$, meaning a ferroelectric to paraelectric transition takes place before the polydomain to monodomain transition. However, $\dc$ is also very sensitive to the environment of the film. For a sandwich system with a thin film of PTO between two regions of STO, again $\dc \gg \dm$. For strongly-coupled FE/PE superlattices (small $\a$), however, $\dm$ increases with $\ks$, and we would have $\dm \gg \dc$, and therefore the thin-limit behaviour presented above should be observable before the films becoming paraelectric.

  The comparative study offered in this work, however, gives the 
expected behaviour of ferroelectric/dielectric heterostructures within the simplest 
Kittel continuum model (continuum electrostatics for a given spontaneous polarization
and dielectric response, plus ideal domain wall formation).
 The described behaviors are already quite rich, and we think they represent a 
paradigmatic reference as basis for the understanding of more complex effects. 
  In particular for superlattices, the strong to weak coupling regime separation based on
this simplest model should be a useful guiding concept.

%In spite of these approximations, the more general theory presented in this paper provides a useful correction to Kittel's law for an isolated ferroelectric thin film, and has allowed us to understand the behaviour of domain structures in the ultrathin limit. This could be useful for fitting experimental measurements of ferroelectric domain structures in thin films, where Kittel's law for an isolated film is known to underestimate the domain widths. The results for FE/PE superlattices in the ultrathin limit are very interesting, and the critical ratio $\acrit$ could provide some insight into the coupling between neighboring layers in experimentally fabricated superlattices.

\section*{Acknowledgments}
The authors would like to thank Pablo Aguado-Puente for helpful discussions. DB would like to acknowledge the EPSRC Centre for Doctoral Training in Computational Methods for Materials Science under grant number EP/L015552/1.

\section*{Appendix A: Electrostatics}
\label{appendix:elec}

\begin{figure}[!ht]
%\hspace*{-1cm}
\centering
\includegraphics[width=\columnwidth]{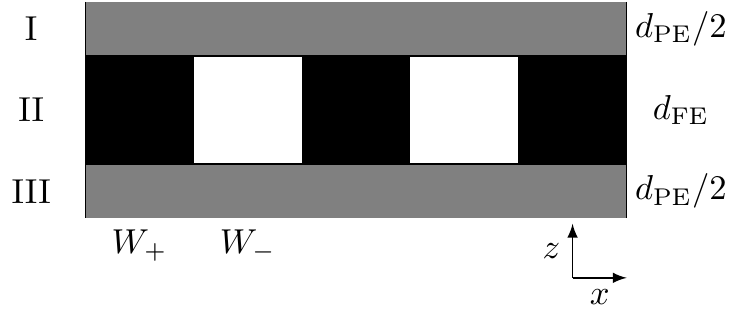}
\caption{The geometry of a FE/PE superlattice. Regions I and III correspond to half of a paraelectric layer each and region II is the ferroelectric layer. The thicknesses of the layers are indicated on the right and $W_+$ and $W_-$ are the widths of the different domain orientations. The black squares are positive domains, with polarization $+P$ and the white squares are negative domains with polarization $-P$. The system is periodic in the horizontal and vertical directions, with periods $W=W_+ + W_-$ and $D=d_{\text{FE}} + d_{\text{PE}}$, respectively}
\label{fig:superlattice_unit_cell}
\end{figure}

Following Ref. [\onlinecite{springer}], we obtained the expressions for the electrostatic energies of the OL, SW and SL systems. We present the derivation for the SL system, but the method also applies to the OL and SW systems, the only difference being that the boundary conditions change from periodic to infinite.

Consider a periodic array of ferroelectric and paraelectric layers as shown in Fig. \ref{fig:superlattice_unit_cell}. The ferroelectric layer has a structure of $180^{\circ}$ stripe domains with polarization $\pm P$ and widths $W_+$, $W_-$. The unit cell of such a system is formed by one positive and one negative polarization domain in the $x$-direction, with period $W=W_+ + W_-$, and one ferroelectric and one paraelectric layer in the $z$-direction, with period $D = d_{\text{FE}}+d_{\text{PE}}$. As mentioned previously, we assume that the widths of the domain walls are infinitely thin. Thus, we write the polarization as a Fourier series:
\beq{superlattice_polarisation}
P(x)= AP+ \sum_{n=1}^{\infty}\frac{4P}{n\pi}\sin{\lb\frac{n\pi}{2}(A+1)\rb}\cos{\lb n kx\rb}
\eec
where $A = \frac{W_+ - W_-}{W}$ is the mismatch between the domains and $k = \frac{2\pi}{W}$. We can see that the polarization is split into a monodomain term, the average polarization $AP$, and polydomain terms in the infinite series. The polydomain limit is obtained when $A\to 0$, i.e. the domain widths are equal. The monodomain limit is obtained when $A\to\pm 1$, i.e. one of the domain widths tends to zero. To obtain the electric fields in the SL, we must first determine the electrostatic potentials. They satisfy the following Laplace equations:
\beq{laplace}
\begin{split}
\kappa_{ij}\partial_{i}\partial_{j}\pII &= 0 \\
\ks\grad\pI = \ks\grad\pIII  & = 0
\end{split}
\eec
where regions I, II and III are the different parts of the unit cell as shown in Fig. \ref{fig:superlattice_unit_cell}. Since the terms in \eqref{superlattice_polarisation} are linearly independent, we can treat the monodomain and polydomain cases separately. Clearly the potentials must be even and periodic in $x$, so the general solutions to \eqref{laplace} are of the form
\beq{potentials2}\resizebox{0.85\columnwidth}{!}{$
\begin{split}
\pI(x,z) &=  c_0^1(z)+\sum_{n=1}^{\infty}\cos{\lb nkx \rb}\lb c_n^1e^{n k z} + d_n^1e^{-n k z}\rb\\
\pII(x,z) &= c_0^2(z)+\sum_{n=1}^{\infty}\cos{\lb nkx \rb}\lb c_n^2e^{nk\sqrt{\frac{\ka}{\kc}}z} + d_n^2e^{-nk\sqrt{\frac{\ka}{\kc}}z}\rb \\
\pIII(x,z) &= c_0^3(z)+\sum_{n=1}^{\infty}\cos{\lb nkx \rb}\lb c_n^3e^{n k z} + d_n^3e^{-n k z}\rb
\end{split}$}
\eep
In order to obtain the potentials, we must use the symmetries and boundary conditions of the system to determine the coefficients:
\beq{}
\begin{split}
\pI (d_{\text{FE}}/2) & = \pII(d_{\text{FE}}/2) \\
\pIII (-d_{\text{FE}}/2) & =\pII(-d_{\text{FE}}/2) \\
\pI (D/2) & =\pIII(-D/2) \\
(\vec{D}_{\text{I}}-\vec{D}_{\text{II}})\cdot\hat{n} & = 0\\
(\vec{D}_{\text{III}}-\vec{D}_{\text{II}})\cdot\hat{n} & = 0\\
\pI(z)& =-\pIII(-z)
\end{split}
\eep
The first two conditions are obtained by matching the potentials at the interfaces. The third comes from imposing periodic boundary conditions on the unit cell. The fourth and fifth are obtained by matching the normal components of the displacement fields,
\beq{d_fields}
\begin{split}
\vec{D}_{\text{I}}& = \eo\ks\vec{E}_{\text{I}} \\
\vec{D}_{\text{II}}& = \eo\kappa \vec{E}_{\text{II}} + \vec{P} \\
\vec{D}_{\text{III}}& = \eo\ks\vec{E}_{\text{III}}
\end{split}
\eec
at the interfaces, and the final condition is obtained from the symmetry of the system under $z\to -z$.

After some algebra, we find that the potentials are given by
\begin{widetext}
\beq{}
\begin{aligned}
\pI (z) & = -\frac{AP_S}{\eo\left[\frac{\kc}{d_{\text{FE}}}+\frac{\ks}{d_{\text{PE}}}\right]d_{\text{PE}}}(z-D/2) - \sum_{n=1}^{\infty}\alpha_n \beta_n \frac{\cos{\lb n k x\rb}\sinh{\lb nk \lb z-D/2\rb\rb}}{\x\kc\cosh{\lb nk\x \frac{d_{\text{FE}}}{2}\rb}+\ks\coth{\lb nk \frac{d_{\text{PE}}}{2}\rb}\sinh{\lb nk\x \frac{d_{\text{FE}}}{2}\rb}}\\
\pII (z)& = \frac{AP_S}{\eo\left[\frac{\kc}{d_{\text{FE}}}+\frac{\ks}{d_{\text{PE}}}\right]d_{\text{FE}}}z + \sum_{n=1}^{\infty}\alpha_n \frac{\cos{\lb n k x\rb}\sinh{\lb nk\x z\rb}}{\x\kc\cosh{\lb nk\x \frac{d_{\text{FE}}}{2}\rb}+\ks\coth{\lb nk \frac{d_{\text{PE}}}{2}\rb}\sinh{\lb nk\x \frac{d_{\text{FE}}}{2}\rb}}\\
\pIII (z)& = -\frac{AP_S}{\eo\left[\frac{\kc}{d_{\text{FE}}}+\frac{\ks}{d_{\text{PE}}}\right]d_{\text{PE}}}(z+D/2)- \sum_{n=1}^{\infty}\alpha_n \beta_n\frac{ \cos{\lb n k x\rb}\sinh{\lb nk\lb z+D/2\rb\rb}}{\x\kc\cosh{\lb nk\x \frac{d_{\text{FE}}}{2}\rb}+\ks\coth{\lb nk \frac{d_{\text{PE}}}{2}\rb}\sinh{\lb nk\x \frac{d_{\text{FE}}}{2}\rb}}
\end{aligned}
\eec
\end{widetext}
where
\beq{}
\begin{split}
\alpha_n &= \frac{4P}{\eo n^2\pi k}\sin{\lb \frac{n\pi}{2}(A+1)\rb} \\
\beta_n &= \frac{\sinh{\lb n k\x \frac{d_{\text{FE}}}{2}\rb}}{\sinh{\lb n k \frac{d_{\text{PE}}}{2}\rb}}
\end{split}
\eep
The monodomain part of the potential has a zig-zag shape as expected, which is sensitive to the ratio of layer thicknesses and permittivities. The electrostatic energy of the system is obtained from
\beq{}
\F = \frac{1}{2}\int \kappa_{ij}E_iE_j\dd{x}\dd{z}
\eec
where the fields are the gradients of the potentials: $\vec{E}=-\vec{\nabla}\p$. We integrate over the domain period in the $x$-direction and over both layers in the $z$-direction. Finally, the total electrostatic energy of the system is given by the second line of Eq. \eqref{full_elec}. The first and third lines are the electrostatic energies of the SW and SL cases respectively, obtained using the same method. In all cases the energy is conveniently split into monodomain and polydomain parts. We can see that the monodomain parts for the OL and SW cases are identical to that of a thin film in a vacuum, as expected. We can also see that the polydomain part vanishes when $A\to\pm 1$, and the polydomain energy is obtained when $A\to 0$.

It will be useful for us to work in terms of energy \textit{per unit volume}. For the OL and SW cases, we simply divide by the thickness the thin film. For the superlattice however, we must use the total volume of the unit cell. For convenience, we would like to work in terms volume of the ferroelectric layer. So we let
\beq{}
\begin{split}
d &= d_{\text{FE}} \\
\alpha &= \frac{d_{\text{PE}}}{d_{\text{FE}}}
\end{split}
\eec
so that
\beq{}
\begin{split}
d_{\text{PE}} &= \alpha d \\
D & = (1+\alpha)d
\end{split}
\eep

The energies in Eq. \eqref{full_elec} give a complete picture of the electrostatics of ferroelectric thin films and superlattices.
\begin{widetext}
\beq{full_elec}
\begin{aligned}
\F_{\text{SW}} &= \frac{P^2}{2\eo\kc}\lb A^2 + \frac{16\kc}{\pi^3}\frac{w}{d}\sum_{n=1}^{\infty}\frac{\sin^2{\lb \frac{n\pi}{2}(A+1)\rb}}{n^3}\frac{1}{\ks+\x\kc\coth{\lb \frac{n\pi}{2}\x\frac{d}{w}\rb}}\rb\\
\F_{\text{SL}} &= \frac{1}{(1+\a)}\frac{P^2}{2\eo\kc}\lb\frac{\kc}{\kc + \a^{-1}\ks}A^2 + \frac{16\kc}{\pi^3}\frac{w}{d}\sum_{n=1}^{\infty}\frac{\sin^2{\lb\frac{n\pi}{2}(A+1)\rb}}{n^3}\frac{1}{\x\kc\coth{\lb \frac{n \pi}{2}\x \frac{d}{w}\rb}+\ks\coth{\lb \frac{n \pi}{2}\a\frac{d}{w}\rb}}\rb
\end{aligned}
\eeq
\end{widetext}

For the substrate case, the energy is given by
\beq{full_sub}\resizebox{0.85\columnwidth}{!}{$
\F_{\text{OL}} = \frac{P^2}{2\eo\kc}\lb A^2 + \frac{8\kc}{\pi^3}\frac{w}{d}\sum_{n=1}^{\infty}\frac{\sin^2{\lb \frac{n\pi}{2}(A+1)\rb}}{n^3}\gamma_n^{-2}\Gamma_n \rb$}
\eeq
where
\beq{}\resizebox{0.85\columnwidth}{!}{$
\begin{split}
\gamma_n &= (\x^2\kc^2+\ks)\sinh{\lb n\pi \x\frac{d}{w}\rb}+\x\kc(1+\ks)\cosh{\lb n\pi \x\frac{d}{w}\rb}\\
\Gamma_n & = (\x^2\kc^2-\ks)(1+\ks)-4\x^2\kc^2(1+\ks)\cosh{\lb n\pi \x\frac{d}{w}\rb} \\
         & + (1+\ks)(3\x^2\kc^2+\ks)\cosh{\lb 2n\pi \x\frac{d}{w}\rb} \\
         & - 4\x\kc(\x^2\kc^2+\ks)\sinh{\lb n\pi \x\frac{d}{w}\rb}\\
         & +\x\kc(1+2\x^2\kc^2+\ks(4+\ks))\sinh{\lb 2n\pi \x\frac{d}{w}\rb}
\end{split}$}
\eeq
It is important to check that the polydomain part of the energy reproduces the monodomain and Kittel energies in the appropriate limits. Letting $A=0$, we have
\beq{}\resizebox{0.85\columnwidth}{!}{$
\F_{\text{SL}} = \frac{P^2}{2\eo\kc}\lb \frac{16\kc}{\pi^3}\frac{w}{d}\sum_{n \ \text{odd}}\frac{1}{n^3}\frac{1}{\x\kc\coth{\lb \frac{n \pi}{2}\x \frac{d}{w}\rb}+\ks\coth{\lb \frac{n \pi}{2}\a\frac{d}{w}\rb}}\rb
$}
\eec
ignoring the prefactor of $(1+\a)^{-1}$. The monodomain limit is realized when $w\to\infty$. Using the expansion $\coth{(a x)} \sim \frac{1}{a x}$ about $x=0$, we get
\beq{}
\begin{split}
\F_{\text{SL}}&\to \frac{P^2}{2\eo(\kc+\a^{-1}\ks)}\frac{8}{\pi^2}\sum_{n \ \text{odd}}\frac{1}{n^2}\\
&=\frac{P^2}{2\eo(\kc+\a^{-1}\ks)}
\end{split}
\eec
since $\sum_{n \ \text{odd}}\frac{1}{n^2} = \frac{\pi^2}{8}$. For the Kittel limit, $\frac{d}{w}\gg 1$. Using $\coth{(x)}\to 1$ for large $x$, we get
\beq{}
\F_{\text{SL}}\to \frac{P^2}{2\eo}\frac{14\zeta(3)}{\pi^3}\frac{1}{\ks+\x\kc}\frac{w}{d}
\eec
where we used $\sum_{n \ \text{odd}}\frac{1}{n^3} = \frac{7\zeta(3)}{8}$.

\section*{Appendix B: Asymptotic Approximation of the Domain Width in the Ultrathin Limit}

Following the method in Ref. [\onlinecite{superlattice_domains_1}], we obtain an approximation to the equilibrium domain width behavior in the ultrathin limit. For the IF system, total energy is approximately
\beq{}
\F_{} \cong \frac{\S}{w} + \frac{8P^2}{\eo\kc\pi^2}\frac{1}{\z}\sum_{n=0}^{\infty}\frac{1}{(2n+1)^3}\tanh{\lb \frac{(2n+1)}{2}\z\rb}
\eec
when $\z = \pi\x\frac{d}{w} \ll 1$. Using
\beq{}\resizebox{0.85\columnwidth}{!}{$
\tanh{\lb \frac{(2n+1)}{2}\z\rb} = \int_0^1 \del_{\lambda}\lb \tanh{\lb \frac{(2n+1)}{2}\z\lambda\rb}\rb\dd{\lambda}
$}
\eec
we get
\beq{}
\begin{split}
\F & \cong  \frac{\S}{w} + \frac{4P^2}{\eo\kc\pi^2}\int_0^1\dd \lambda\sum_{n=0}^{\infty}\frac{1}{(2n+1)^2}\frac{1}{\cosh^2{\lb \frac{(2n+1)}{2}\z\lambda\rb}}\\
&\approx \frac{\S}{w} + \frac{16P^2}{\eo\kc\pi^2}\int_0^1\dd \lambda\sum_{n=0}^{\infty}\frac{e^{-(2n+1)\z\lambda}}{(2n+1)^2}
\end{split}
\eep

From Ref. [\onlinecite{superlattice_domains_1}]: 
\beq{}
\int_0^1\dd \lambda\sum_{n=0}^{\infty}\frac{e^{-(2n+1)\z\lambda}}{(2n+1)^2} = \frac{\pi^2}{8}-\frac{\z}{4}\ln{\lb\frac{e^p}{\z}\rb}+\bigO(\z^3)
\eec
where $p=\frac{1}{2}(3+\ln{(4)})$. Thus, our approximation to the energy becomes
\beq{}
\F \cong \frac{\S}{w}+\frac{P^2}{2\eo\kc}+ \frac{P^2}{2\eo\kc}\lb3-\frac{8}{\pi}\x\frac{d}{w}\ln{\lb\Lambda\frac{w}{d}\rb}\rb
\eec
where
\beq{}
\Lambda = \frac{e^p}{\pi\x}
\eep
The first two terms are the domain energy and monodomain energy, and the third term is an asymptotic correction. Minimizing with respect to $w$, we get 
\beq{}
w(d) = \frac{\pi\x}{2\sqrt{e}}d\exp{\lb\frac{\pi^2}{8}\frac{\kc}{\x}\b\frac{l_k}{d}\rb}
\eep
The corresponding minimum width is
\beq{}
 \dm = \frac{\pi^2}{8}\frac{\kc}{\x}\b l_k
\eep

\section*{Appendix C: Analytic Approximation to the Domain Width}

We can obtain an analytic approximation to the equilibrium domain behavior if we replace the electrostatic energy with a simpler function which reproduces the monodomain and Kittel energies in the appropriate limits. For the IF system, we could use:
\beq{}
\F_{\text{e}}^* = \underbrace{\frac{P^2}{2\eo\kc}}_{\F_{\text{mono}}}\frac{1}{1+\frac{1}{\kc\b}\frac{d}{w}}
\eep
When $w/d$ is very large, the second term in the denominator goes to zero and we get $\F_{\text{e}}^* = \F_{\text{mono}}$. When $w/d$ is very small, the second term in the denominator dominates and we get $\F_{\text{elec}}^* = \frac{P^2}{2\eo}\b\frac{w}{d} = \F_{\text{Kittel}}$. This approximation can also be used for the OL and SW systems, since the extension to these systems is simply achieved via $\b \to \b(\ks)$. For the superlattice, the energy in the monodomain limit is different:
\beq{}
\F_{\text{mono,SL}} = \frac{1}{(1+\a)}\frac{P^2}{2\eo(\kc+\a^{-1}\ks)}
\eeq
The prefactor $(1+\a)^{-1}$ scales the energy with the ratio of the layer thicknesses. The energy cost of creating a domain structure is scaled by this prefactor. Thus, the equilibrium domain width will be unaffected by this prefactor, and we can neglect it. Now, the monodomain energy for a SL is similar to the case of a thin film, but with renormalized permittivity: $\kc\to\kc + \a^{-1}\ks$. When $\a\to\infty$, the thin film expressions are recovered, so we can work with the SL system and the other systems can be recovered by taking $\a\to\infty$ and the correct choice of $\b(\ks)$.

The total energy for the SL system is
\beq{F_approx_SL}
\F_{\text{SL}}^*  = \frac{\S}{w}+\frac{F_{\text{mono,SL}}}{1+\frac{x}{w}}
\eec
where
\beq{}
x = \frac{d}{(\kc+\a^{-1}\ks)\b(\ks)}
\eep
Minimizing Eq. \eqref{F_approx_SL}, we get
\beq{w}
w(d) = \frac{\sqrt{l_k(\ks)d}}{1-(\kc+\a^{-1}\ks)\b(\ks)\sqrt{\frac{l_k(\ks)}{d}}}
\eep
Clearly, this expression has square root behavior for large $d$ (Kittel) and diverges for small $d$ (monodomain). The width diverges at
\beq{}
\dinf = (\kc+\a^{-1}\ks)^2\b(\ks)^2l_k(\ks)
\eec
and has a minimum at
\beq{}
\begin{split}
\dm &= 4(\kc+\a^{-1}\ks)^2\b(\ks)^2l_k(\ks) = 4\dinf\\
 & = 8\eo(\kc+\a^{-1}\ks)^2\b(\ks)\S\frac{1}{P^2}
\end{split}
\eep

Interestingly, the relation $\dm = 4\dinf$ is universal and independent of system-specific parameters.

%\begin{widetext}
%\beq{}\resizebox{0.9\columnwidth}{!}{$
%\F_{\text{Sub}} = \frac{P^2}{2\eo\kc}\lb A^2 + \frac{8\kc}{\pi^3}\frac{w}{d}\sum_{n=1}^{\infty}\frac{\sin^2{\lb \frac{n\pi}{2}(A+1)\rb}}{n^3}\frac{(\ka\kc-\ks)(1+\ks)-4\ka\kc(1+\ks)\cosh{\lb n\pi \sqrt{\frac{\ka}{\kc}}\frac{d}{w}\rb}+(1+\ks)(3\ka\kc+\ks)\cosh{\lb 2n\pi \sqrt{\frac{\ka}{\kc}}\frac{d}{w}\rb} - 4\sqrt{\ka\kc}(\ka\kc+\ks)\sinh{\lb n\pi \sqrt{\frac{\ka}{\kc}}\frac{d}{w}\rb}+\sqrt{\ka\kc}(1+2\ka\kc+\ks(4+\ks))\sinh{\lb 2n\pi \sqrt{\frac{\ka}{\kc}}\frac{d}{w}\rb}}{\lb(\ka\kc+\ks)\sinh{\lb n\pi \sqrt{\frac{\ka}{\kc}}\frac{d}{w}\rb}+\sqrt{\ka\kc}(1+\ks)\lb\cosh{\lb n\pi \sqrt{\frac{\ka}{\kc}}\frac{d}{w}\rb}\rb\rb^2}\rb$}
%\eeq
%\end{widetext}

%%% REFERENCES %%%

%\bibliographystyle{apsrev4-1} % Tell bibtex which bibliography style to use
%\bibliography{references.bib}


\end{document}